# Rational design of large anomalous Nernst effect in Dirac semimetals

Panshuo Wang[1], Zongxiang Hu[1], Xiaosong Wu[2] and Qihang Liu[1,3,4,*]

[1]*Shenzhen Institute for Quantum Science and Engineering and Department of Physics, Southern University of Science and Technology, Shenzhen 518055, China*

[2]*State Key Laboratory for Artificial Microstructure and Mesoscopic Physics, Frontiers Science Center for Nano-optoelectronics, and Collaborative Innovation Center of Quantum Matter, Peking University, Beijing, China*

[3]*Guangdong Provincial Key Laboratory of Computational Science and Material Design, Southern University of Science and Technology, Shenzhen 518055, China*

[4]*Shenzhen Key Laboratory of Advanced Quantum Functional Materials and Devices, Southern University of Science and Technology, Shenzhen 518055, China*

[*]Email: liuqh@sustech.edu.cn

## Abstract

Anomalous Nernst effect generates a transverse voltage perpendicular to the temperature gradient. It has several advantages compared with the longitudinal thermoelectricity for energy conversion, such as decoupling of electronic and thermal transports, higher flexibility, and simpler lateral structure. However, a design principle beyond specific materials systems for obtaining a large anomalous Nernst conductivity (ANC) is still absent. In this work, we theoretically demonstrate that a pair of Dirac nodes under a Zeeman field manifests a double-peak anomalous Hall conductivity curve with respect to the chemical potential and a compensated carriers feature, leading to an enhanced ANC pinning at the Fermi level compared with that of a simple Weyl semimetal with two Weyl nodes. Based on first-principles calculations, we then provide two Dirac semimetal candidates, i.e., $Na_3Bi$ and NaTeAu, and show that under a Zeeman field they exhibit a sizable ANC value of 0.4 $A/(m \cdot K)$ and 1.3 $A/(m \cdot K)$, respectively, near the Fermi level. Our work provides a design principle with a prototype band structure for enhanced ANC pinning at Fermi level, shedding light on the inverse design of other specific functional materials base on electronic structure.

**Introduction**

Anomalous Nernst effect (ANE), for which the generated voltage drop is perpendicular to the temperature gradient without any assistance of external magnetic field, attracts sustaining attentions for energy harvesting [1-5]. The nature of transverse transport helps for device miniaturization during the energy conversion. Compared with the ordinary Nernst effect where the carriers are driven by Lorentz force provided by the external magnetic field [6], the driving force of ANE can be classified as extrinsic and intrinsic contributions. While the extrinsic contributions refer to the drags of magnon or the skew scattering and side jump due to strong spin-orbit coupling [7-8], the intrinsic one is attributed to the effects of Berry curvature of the energy bands [9-10]. Berry curvature behaves as the magnetic field deflecting the electrons in momentum space, diverging at the band crossing points such as Weyl points or nodal lines. Therefore, the corresponding topological semimetals are favorable to exhibit large anomalous Hall conductivity (AHC) [11-17]. Similarly, the recent studies pursuing materials candidates with large ANE also focus on topological semimetals. For example, sizable anomalous Nernst conductivity (ANC) [~ 0.1-10 A/(mK)] is observed in topological materials $Co_2MnGa$ [18], $Co_3Sn_2S_2$ [19], $Mn_3X$ (X = Sn, Ge) [20], and Heusler family [21-22], all of which possess multiple Weyl points or nodal lines. However, the ANC $\alpha$ and the corresponding thermopower $S$ are still too small for practical applications compared with magneto-thermoelectricity [23].

The Mott relation [4,24] reflects the connection between the AHC σ and ANC α, both of which are related to the integration of the Berry curvature through the Brillouin zone. At the low temperature limit, the Mott relation expresses as:

$$\alpha_{ij} = \frac{\pi^2}{3}\frac{k_B^2 T}{e}\sigma'_{ij}(\varepsilon_F), \qquad (1)$$

where $k_B$, $T$, $e$ and $\varepsilon_F$ are Boltzmann constant, temperature, elementary charge, and Fermi energy, respectively. The Mott relation indicates that the maximum value of ANC α occurs at the energy with the steepest slope of AHC σ. Thus, a band structure with opposite AHC peaks on the opposite side of the Fermi level $\varepsilon_F$ is expected to host large ANC. In addition, the ANC can be further enhanced if carriers are compensated at the current Fermi level. Although carriers of opposite charge play against each other in the Seebeck effect [25], they synergically contribute to the ordinary Nernst effect, giving rise to a giant and non-saturating Nernst voltage [26-28]. In order to utilize the similar mechanism in ANE, it is necessary to design a system in which both types of carriers experience a fictitious magnetic field, i.e., Berry curvature, of the same sign. However, Among the various materials with ANE observed in experiments, they are almost discovered by chance or in some special material systems, e.g., Heusler compounds [21-22]. Moreover, for most theoretically proposed ANE materials, the maximum value of ANC $\alpha$, corresponding to a specific chemical potential, usually deviates from the pristine Fermi level. It means that to achieve the best performance the Fermi level needs to be tuned to the optimal ANC through electron or hole doping. Therefore, it is desirable to conduct a material design from fundamental electronic structure that hosts large ANE pinning at the intrinsic Fermi level.

In this work, rather than resorting to a single material system, we theoretically illustrate that the target functionality of enhanced ANC pinning at Fermi level can be treated as an electronic-structure design problem. Starting from a model Hamiltonian of a Dirac semimetal under Zeeman field, we reveal two pairs of Weyl points forming a "G-type" configuration of chirality in momentum space, leading to a double-peak feature of the AHC. Such a system also features compensated carriers, both types of which experience Berry curvatures of the same sign. As a result, a 300% enhanced ANC pinning at Fermi level is obtained compared with a simple two-band Weyl semimetal model. Exemplified by two realistic Dirac materials $Na_3Bi$ [29-30] and NaTeAu [31-32], we then perform ANC calculations based on atomistic Hamiltonians obtained from density-functional theory and obtained sizable ANC in the vicinity of the Fermi level (~0.38 $A/(m \cdot K)$ and ~1.3 $A/(m \cdot K$, respectively). Our work paves an avenue for the material realization of large ANE materials.

**Double-peak AHC of a Dirac semimetal under Zeeman field**

A Weyl point is the band crossing position in momentum space, where the Berry curvature diverges, identified as the source or sink of Berry curvature [10]. It always connects with the large AHC and ANC. First, we study the electronic structure and properties of a tilted two-band Weyl model system. The minimal Hamiltonian for such a model can be written as [33]:

$$H = A(k_x\sigma_x + k_y\sigma_y) + M(k_w^2 - \boldsymbol{k}^2)\sigma_z + tk_z, \qquad (2)$$

where $\sigma$ are Pauli matrices, and $A$, $M$, $k_w$ and $t$ are model parameters. $A$ and $M$ are relating to the Fermi velocity and the inverse of effective mass near the band edge, respectively. $k_w$ characterizes the distance of the two Weyl points in momentum space and $t$ describes the tilting of the band structure along $k_z$. This minimal three-dimensional model describes a pair of Weyl nodes locating at $(0,0,\pm k_w)$, with an energy difference of $2tk_w$. The monopole charges, i.e., chirality of Weyl nodes at $(0,0,\pm k_w)$, are $\mp 1$. By adopting a set of parameters that $A = 0.18\ eV \cdot Å$, $M = 2\ eV \cdot Å^2$, $k_w = sqrt(0.1)\ Å^{-1}$ and $t = 0.3\ eV \cdot Å$, figure 1a displays the band structure of the tilted two-band Weyl model. The anomalous Hall conductivity is an even function with respect to the chemical potential (see Fig 1b), with one peak ($\sim 220\ 1/(\Omega \cdot cm)$) locating at $E_f = 0\ eV$. The corresponding ANC curve is an odd function with respect to the chemical potential, consisting with the Mott relation. The two opposite peaks of ANC connect with the maximum slope of AHC curve symmetrically locating above and below the Fermi level with the absolute value of about $1.1 A/(m \cdot K)$ (see Fig 1c). Such result is similar to the two-band Weyl model with no band tilt [34]. In a word, a pair of Weyl points with opposite chirality generates only one AHC peak at Fermi level. Such a single-peak Weyl model gives rise to a moderate ANE away from the Fermi level, while the ANE at the fermi level is exactly zero.

To gain an enhanced ANC pinning at Fermi level, we require the AHC curve is an odd function with two opposite peaks locating above and below the Fermi level within a narrow energy interval. This requires at least four Weyl nodes with one pair locating below the Fermi level and the other above the Fermi level possessing a "G-type" arrangement of chirality. Inspired by the results of two-band Weyl model, we adopt a Dirac model, i.e., a two-band Weyl model connecting with its time-reversal partner, combining with a Zeeman term breaking the time-reversal symmetry. Such a four-band Hamiltonian is written as [35]:

$$H = \begin{bmatrix} (k_w^2 - \boldsymbol{k}^2)M & Ak_x - iAk_y & 0 & 0 \\ Ak_x + iAk_y & -(k_w^2 - \boldsymbol{k}^2)M & 0 & 0 \\ 0 & 0 & (k_w^2 - \boldsymbol{k}^2)M & -Ak_x - iAk_y \\ 0 & 0 & -Ak_x + iAk_y & -(k_w^2 - \boldsymbol{k}^2)M \end{bmatrix} + \begin{bmatrix} S & 0 & 0 & 0 \\ 0 & S & 0 & 0 \\ 0 & 0 & -S & 0 \\ 0 & 0 & 0 & -S \end{bmatrix},$$

(3)

where $A$, $M$ and $k_w$ are parameters as that in Eq. (2), and $S$ denotes the strength of Zeeman term. Consisted by two copies for different spin degree of freedom, the Hamiltonian can be written in a concise form:

$$H = A(k_x \sigma_z \otimes \sigma_x + k_y \sigma_0 \otimes \sigma_y) + M(k_w^2 - \boldsymbol{k}^2)\sigma_0 \otimes \sigma_z + S\sigma_z \otimes \sigma_0. \quad (4)$$

Without the Zeeman term in Eq. (3), the energy bands are doubly degenerate, with two Dirac nodes locating at $(0, 0, \pm k_w)$. Each Dirac node contains two degenerate Weyl nodes with opposite chirality. Due to the time-reversal symmetry, both the AHC and ANC are vanishing. However, the compensation of the Weyl points for opposite spin channel ensures that when the time-reversal symmetry is slightly breaking, there could be significant response for AHC and ANC. As shown in Fig. 1(d), by adopting the same parameters as that of the tilted Weyl model in Eq. (2) and a Zeeman term $S = 0.05\ eV$, the four Weyl points emerge, with the chiralities arranging a G-type configuration around the Fermi level (see Fig. 1(d)). The anomalous Hall conductivity is an odd function with respect to the chemical potential (see Fig. 1e), with two opposite peaks ($\sim 219\ 1/(\Omega \cdot cm)$) locating symmetrically above and below the Fermi level.

The corresponding ANC curve exhibits an even function with respect to the chemical potential, in consistent with the Mott relation. Instead of being zero at the Fermi level in the single-peak Weyl model, the ANC curve in the two-peak Dirac model shows the maximum value right at the Fermi level, with the magnitude $3.2 A/(m \cdot K)$ almost 300% of the maximum value for the Weyl model (see Figs. 1c and 1f). Moreover, the two smaller symmetric peaks ($\sim 1.3 A/(m \cdot K)$) locating on both sides of the Fermi level are also comparable to the maxima of the Weyl model. All in all, the minimal four-band Dirac model with a Zeeman term displays two peaks of AHC with an enhanced ANC right at the Fermi level. This provides a prototype band feature for designing new ANC real materials.

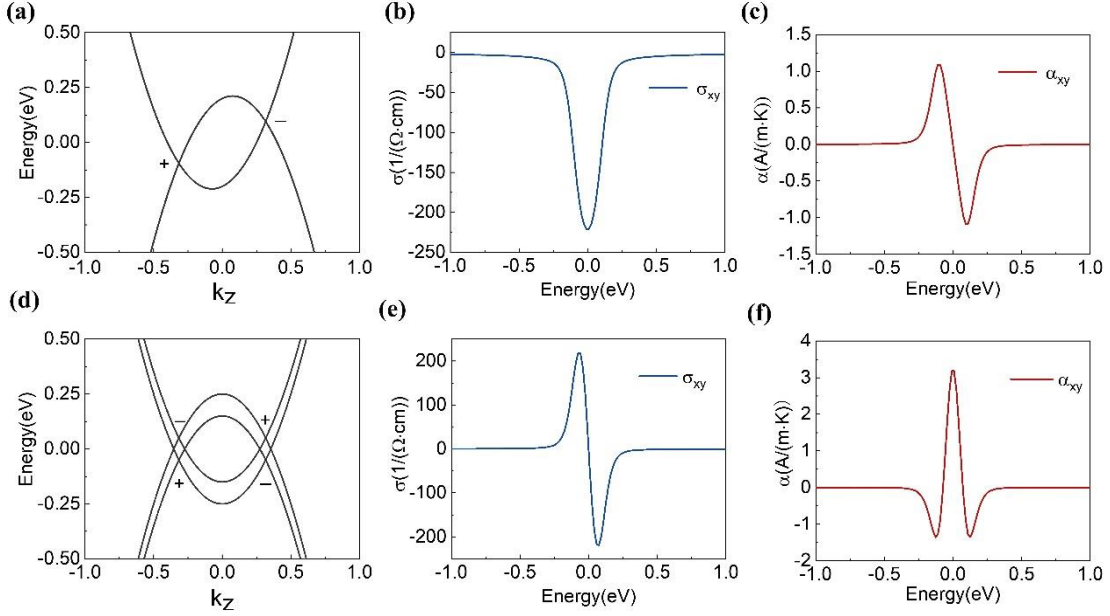

Fig. 1. (a) Band structure of the tilted Weyl semimetal and its (b) AHC and (c) ANC with respect to the chemical potential. (d) Band structure of the four-band Dirac model with Zeeman term and its (e) AHC and (f) ANC with respect to the chemical potential. The ANC curve in (f) behaves as an even function, with the maximum value of about $3.2 A/(m \cdot K)$, which is almost 300% enhancement compared with that of the two-band Weyl model (c). The signs "$\pm$" in the band structures (a) and (d) denote the chirality of Weyl points.

We next consider the modulation of ANC for the four-band Dirac model with Zeeman term as a function of different physical parameters in Eq. (3). Here $A$, $M$, $k_w$ and $S$ relate to the Fermi velocity, inverse of the effective mass of the band edge, distance of the two Dirac nodes in momentum space and strength of the Zeeman field, respectively. In real materials, these physical parameters can be regulated by the strain, element substitution, magnetic proximity effect and external magnetic field etc. Fig. 2 shows the variation of the maximum peak at $E_f = 0 eV$ of ANC (see Fig. 1f) with respect to the four physical parameters in Eq. (3) at different temperatures. First, we study the ANC variations at room temperature (black lines). The peak value of ANC first increases and then decreases with the Fermi velocity $A$ with an optimal value (~ 0.16 $eV \cdot Å$ ) in the present parameter space (see Fig. 2a). The Fermi velocity $A$ describes the slope of the energy band around Dirac points. The maximum value of ANC saturates with the increase of the inverse of the effective mass $M$, i.e., decrease of the effective mass near the band edge (see Fig. 2b). The energy band becomes flatter with a smaller effective mass near the band edge, leading to a larger Berry curvature due to the smaller energy gaps and finally a larger ANC. With the increase of the distance of the two Dirac nodes $k_w$, the peak of ANC gradually increases as shown in Fig. 2c. The Zeeman field strength relates to the energy splitting of the two spin channels (Fig. 1d) and the energy interval of the opposite peaks of AHC (Fig. 1e). A larger Zeeman field strength $S$ represents a stronger breaking of the time-reversal symmetry and hence a larger ANC, while a larger energy interval of the two peaks of AHC will lead to a smaller slope of AHC and suppresses the ANC. This competing mechanism makes the maximum value of ANC increase firstly and then decrease, with an optimal value of about 60meV (Fig. 2d). At 200 K, the ANCs vary similarly to that of the room temperature. Due to the temperature broadening effect, ANCs of a relative low temperature (i. e., 100 K) are obviously smaller than that of the high temperature. It is worth noting that the Mott relation is applicable for a large temperature range (i.e., up to 300 K) that the maximum of ANC $\alpha$ is corresponding to the steepest of AHC $\sigma$. The dependence of physical parameters for the peak value of ANC here offers a reference for the real material regulation.

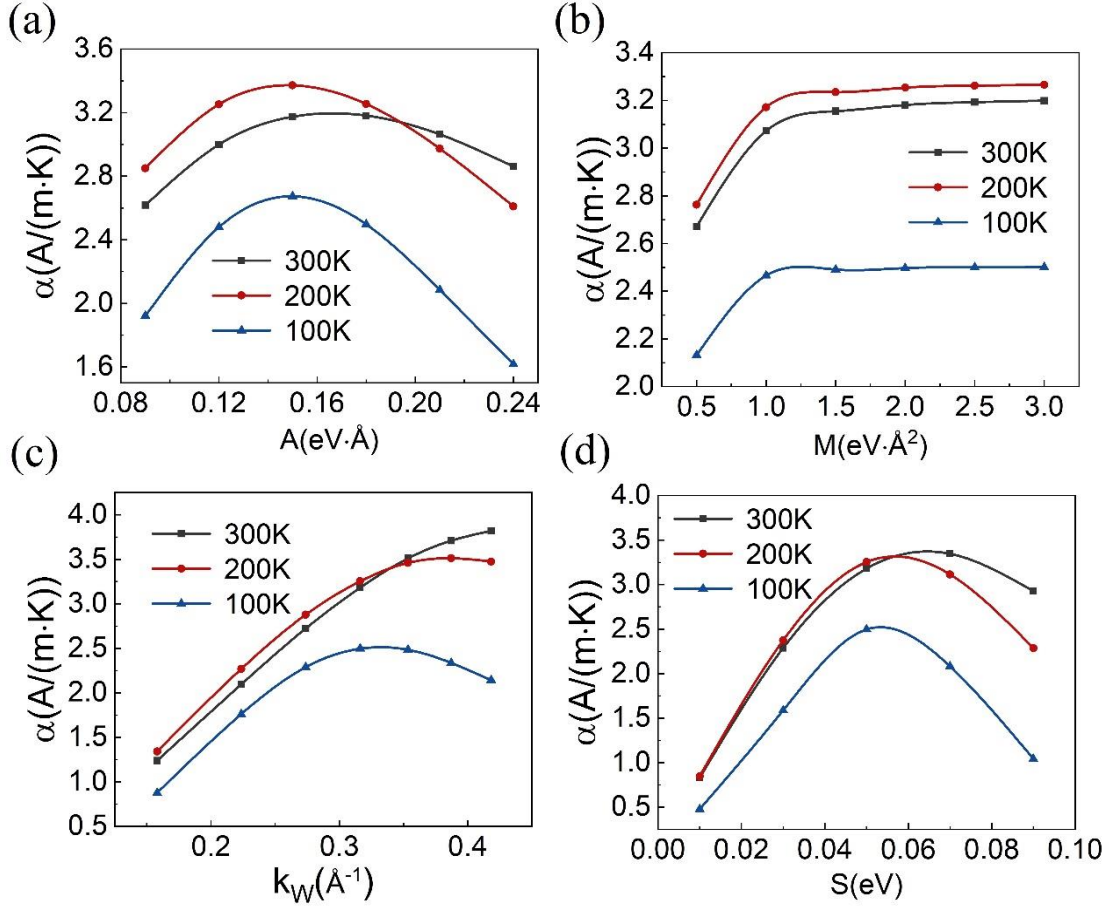

Fig. 2. Variation of the maximum peak at $E_f = 0 eV$ of ANC (see Fig. 1f) with respect to the four physical parameters in Eq. (3). ANC vary with the physical parameters (a) Fermi velocity (A), (b) Inverse of the effective mass around band edge (M), (c) Distance of the two Dirac nodes ($k_w$) and (d) Strength of the Zeeman field (*S*) in Eq. (3).

**Candidate Dirac materials with enhanced ANC**

Exemplified by two Dirac semimetals, we next perform ANC calculations based on atomistic Hamiltonians obtained from density-functional theory to further illustrate the role of double-peak AHC feature. Alkali pnictides Na$_3$Bi crystallizing in hexagonal $P6_3/mmc$ is a widely studied three-dimensional Dirac semimetal [29-30]. With spin-orbit coupling (SOC), the low-energy states form massless Dirac fermions along the $\Gamma - A$ line around Fermi level, as shown in Fig. 3a. The Dirac points are accidental degeneracy protected by the C$_3$ rotational symmetry. By adopting the Na-3s and Bi-6p states around Fermi energy, we construct a tight-binding (TB) model Hamiltonian $H_{TB}$ with Wannier orbitals [36-37] and add a Zeeman term shown as follows:

$$H = H_{TB} + g_e \boldsymbol{\mu} \cdot \boldsymbol{B}, \tag{5}$$

where $g_e$, $\boldsymbol{\mu}$ and $\boldsymbol{B}$ are effective g-factor, magnetic moment and magnetic field, respectively. Fig. 3b shows the band structure under a Zeeman field of 23 meV along the $z$ axis. The broken time-reversal symmetry split the two-fold degenerate bands along the $\Gamma - A$ line, while the branch with a lower Fermi velocity underwent a larger split than the steeper branches.

Due to the absence of time-reversal symmetry, the AHC ($\sigma_{xy}$) of Na$_3$Bi under a magnetic field shows a single maximum peak (~37 $1/(\Omega \cdot cm)$) near the Fermi level (see Fig. 3c), which is similar to the two-band Weyl model in Fig. 1b. As a result, the corresponding ANC of $\alpha_{xy}$ exhibits two opposite peaks (~0.38 $A/(m \cdot K)$) in the vicinity of the Fermi energy (see Fig. 3d). The curve of $\alpha_{xy}$ with respect to the chemical potential is approximately equal to the derivative of $\sigma_{xy}$, in consistent with the Mott relation of Eq. (4). On the other hand, since the Zeeman field here is only along the $z$ axis, the $yz$ and $zx$ components of AHC and ANC are almost negligible compared with that of the $xy$ components.

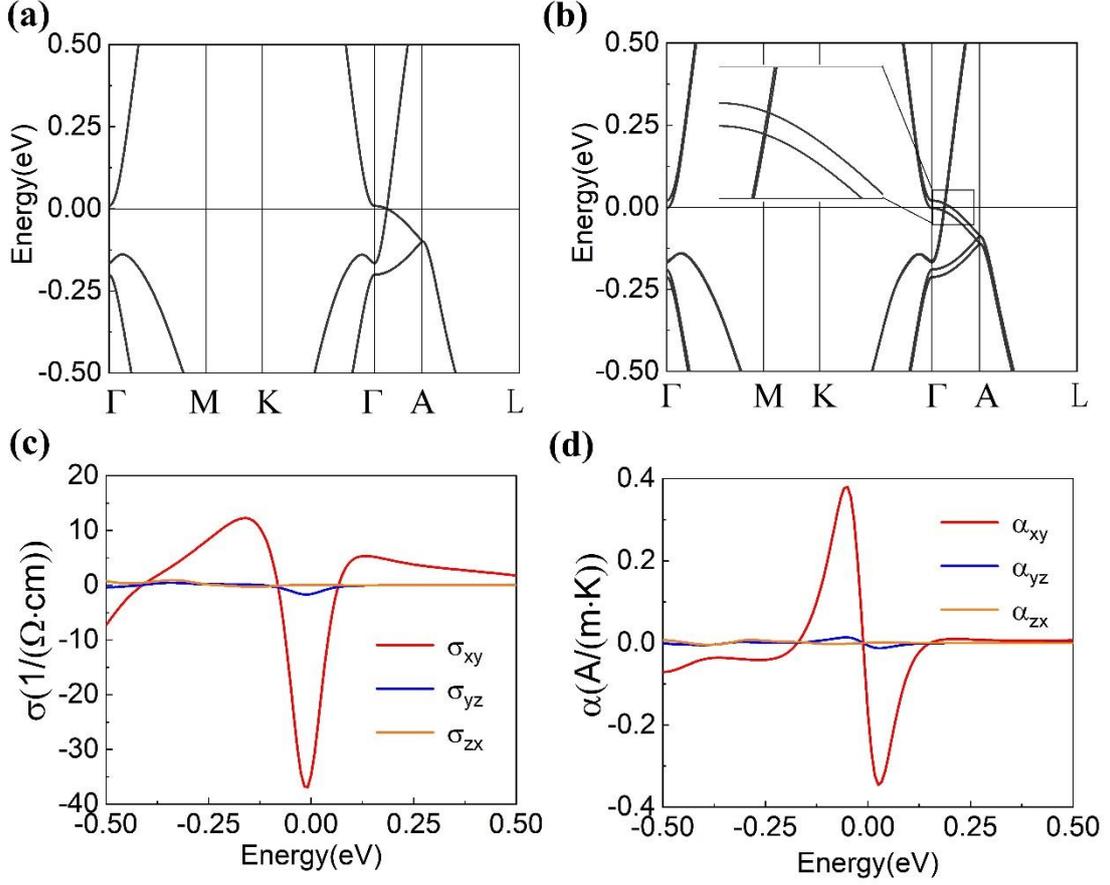

Fig. 3. (a) Band structure of Na₃Bi with the 3D Dirac points along the $\Gamma - A$ line. (b) Band structure of Na₃Bi with Zeeman field. (c) AHC and (d) ANC of Na₃Bi with Zeeman field with respect to the chemical potential.

As the steeper branches forming the Dirac nodes of Na₃Bi hardly split under a Zeeman field, the feature of its AHC displays one peak as that of the two-band Weyl model. And we need a candidate with the Dirac nodes closer to type-I. In the following, the Dirac semimetal NaTeAu with a double-peak AHC is studied. Ternary compound NaTeAu crystalizes in the honeycomb lattice with $P6_3/mmc$ space group [31-32]. Te and Au atoms display a honeycomb lattice positioned on the A and B sites, respectively. The Na atoms form a triangular lattice, stacking alternately with the honeycomb lattice along the c direction. Fig. 4a shows the Dirac nodes of NaTeAu locating on the $\Gamma - A$ line, protected by the $C_3$ rotation symmetry. The two branches forming the Dirac point are very symmetrical, show the type-I feature. By adopting the Na-3s Te-5s, 5p and Au-5d, 6s states around Fermi energy, we construct an effective Hamiltonian $H$ for NaTeAu with $H_{TB}$ and a Zeeman term as shown in Eq. (5). Fig. 4b shows the band structure under a Zeeman field of 29 meV along the $z$ axis. A set of Weyl points emerge under Zeeman field. The AHC ($\sigma_{xy}$) shows a double-peak feature with a sharp slope near the Fermi level (see Fig. 4c). The maximum value of $\sigma_{xy}$ is about 81 $(\Omega \cdot cm)^{-1}$. The ANC curve of $\alpha_{xy}$ exhibits one peak (~1.3 $A/(m \cdot K)$) corresponding the maximum slope of $\sigma_{xy}$ (see Fig. 4d), similar to the feature of minimal four-band Dirac model. On the other hand, the $yz$ and $zx$ components of AHC and ANC are almost zero compared with that of the $xy$ components, since the Zeeman field here is only along the $z$ axis.

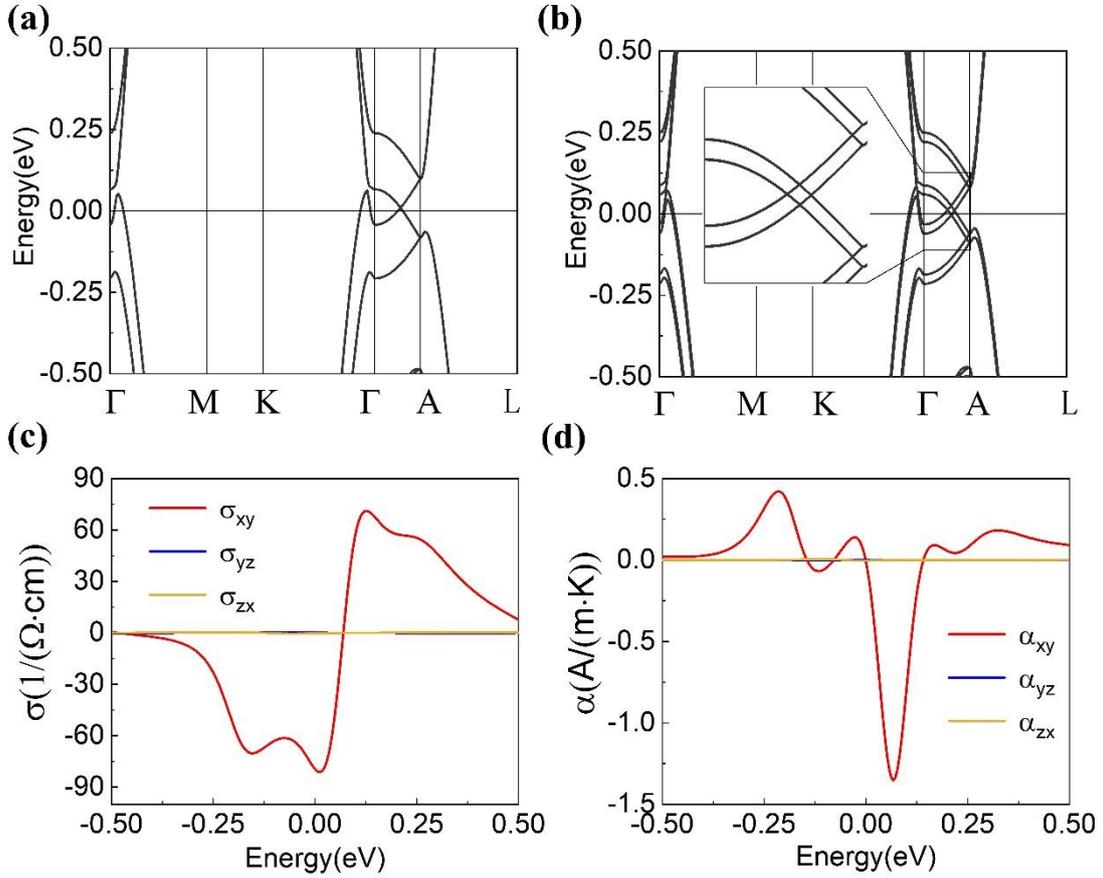

Fig. 4 (a) Band structure of NaTeAu with the 3D Dirac points along the $\Gamma - A$ line. (b) Band structure of NaTeAu with Zeeman field. (c) AHC and (d) ANC of NaTeAu with Zeeman field with respect to the chemical potential.

**Conclusion**

In summary, we first studied the AHC and ANC of the tilted two-band Weyl model.

A pair of Weyl nodes generate one AHC peak. Inspired by such fact, we confirmed that a Dirac $\boldsymbol{k} \cdot \boldsymbol{p}$ model with a Zeeman term will exhibits a two-peak AHC feature and compensated carriers experiencing the Berry curvature of the same sign, leading to about 300% enhanced ANC pinning at Fermi level compared to the two-band Weyl model. Then we studied the regulation routes based on this model. The typical Dirac semimetal Na3Bi with Zeeman field displays two symmetrical peaks near Fermi level, with the maximum value of about 0.38 $A/(m \cdot K)$. Then we predicted that the Dirac semimetal NaTeAu behaves two-peak AHC feature with Zeeman field, possessing a maximum ANC peak of about 1.3 $A/(m \cdot K)$ near Fermi level. Our work provides a design guideline and a protype band structure with enhanced ANC pinning at Fermi level. This will promote the search of new large ANC system and the inverse design of other specific functional materials base on electronic structure.

**Methods**

The anomalous Hall conductivity (AHC) $\sigma$, evaluated via the Kubo formula is formulated as:

$$\sigma_{ij} = \frac{e^2}{\hbar} \sum_n \int \frac{d^3k}{(2\pi)^3} \Omega_{ij}^n f_n, \qquad (6)$$

where $ij$, $n$, $\Omega$, and $f$ are direction indicator, band index, Berry curvature and Fermi-Dirac distribution function, respectively. The anomalous Nernst conductivity (ANC) proposed by Xiao et al. [9-10] can be obtained by substituting the Fermi-Dirac distribution function $f$ by a Gaussian like function $w$, that is:

$$\alpha_{ij} = \frac{e^2}{\hbar} \sum_n \int \frac{d^3k}{(2\pi)^3} \Omega_{ij}^n w_n. \qquad (7)$$

Here $w$ is a function of the temperature $T$:

$$w_n = -\frac{1}{eT}[(\varepsilon_n - \mu)f_n + k_B T \ln(1 + exp\frac{\varepsilon_n - \mu}{-k_B T})], \qquad (8)$$

where $\varepsilon_n$ and $\mu$ are the energy of $n$-th band and the chemical potential, respectively. Comparing the Fermi-Dirac distribution function $f$ and Gaussian like function $w$ with respect to the chemical potential $\mu$, the Gaussian like function $w$ is approximate the derivative of the Fermi-Dirac distribution (see Fig. S1). At low temperature this intrinsic relation explicitly reduces to

$$\alpha_{ij} = \frac{\pi^2}{3}\frac{k_B^2 T}{e}\sigma'_{ij}(\varepsilon_F), \tag{9}$$

which is known as the Mott relation [4,24]. That is to say that the extreme value of ANC $\alpha$ occurs at the maximum slope of AHC $\sigma$. And a band structure with opposite AHCs on either side of the Fermi level $\varepsilon_F$ is expected. At a relative high temperature, such guide rule to obtain large ANC $\alpha$ is also applicable, although the position of the maximum ANC $\alpha$ will slightly shift.

The Berry curvature $\Omega$ adopted in Eq. (1) and Eq. (2) is evaluated via the Kubo formula [10,38]:

$$\Omega_{ij}^n = i\sum_{m\neq n}\frac{\langle n|\frac{\partial H}{\partial k_i}|m\rangle\langle m|\frac{\partial H}{\partial k_j}|n\rangle - (i\leftrightarrow j)}{(E_n-E_m)^2}, \tag{10}$$

where $|n\rangle$ and $E_n$ are the eigenvector and eigenvalue of the H. The Berry curvature acts as the magnetic field in the momentum space, deflecting the carrier transverse to the temperature gradient. The Berry curvature will diverge at the band crossing point, i.e., $E_n = E_m$, what is the Weyl point or nodal line in topological semimetals. In the next section, we will introduce how to construct an efficient Weyl configuration to enhance the ANE.

Electronic structure calculations for real materials were based on the density functional theory (DFT) [39] implemented in the Vienna ab initio simulation package (VASP) [40-41], where the exchange-correlation potential was treated by generalized gradient approximation (GGA) of the Perdew−Burke−Ernzerhof (PBE) functional [42] and the ionic potential was based on the projector augmented wave (PAW) method [43-44]. Owing to the strong relativistic effect in Bi, Te and Au, spin-orbit coupling (SOC) was also considered for the energy band dispersion calculations. The tight-binding model Hamiltonian adapted for the Wannier interpolation of AHC calculations implemented in the WannierTools package [36-37] (i.e., electrical conductivity calculation) was constructed by the Wannier90 software [45] using the maximally localized Wannier function approach [46-48]. For ANC calculations, a subroutine was added to WannierTools package by substituting the Fermi-Dirac function $f$ in Eq. (1)

by the $w$ function of Eq. (3). The s orbits of Na, s, p orbits of Bi and Te, and s, d orbits of Au were selected as the initial projectors for Wannier90 software.

**Acknowledgements**

This work was supported by National Key R&D Program of China under Grant No. 2019YFA0704900, Guangdong Provincial Key Laboratory for Computational Science and Material Design under Grant No. 2019B030301001, the Science, Technology and Innovation Commission of Shenzhen Municipality (No. ZDSYS20190902092905285) and Center for Computational Science and Engineering of Southern University of Science and Technology. X. S. Wu acknowledges financial support by National Key R&D Program of China under Grant No. 2022YFA1403700 and NSFC under Grant No. 12074009.